\documentclass[12pt]{article}
\usepackage{amstext}
\usepackage{amsfonts}
\usepackage{amsmath}
\usepackage{amsthm}
\usepackage{caption}
\usepackage{epsfig}
\usepackage{enumitem}
\usepackage{float}
\usepackage{gensymb}
\usepackage{graphics}
\usepackage{graphicx}
\usepackage{lipsum}
\usepackage{longtable}
\usepackage{lscape}
\usepackage{mathtools}
\usepackage{mathrsfs}
\usepackage{natbib}
\usepackage{pdflscape}
\usepackage{setspace}
\usepackage{times}
\usepackage{titlesec}
\usepackage{verbatim}
\usepackage{fancyhdr}
\usepackage{psfrag,epsf}
\usepackage{enumerate}
\usepackage{url} 

\newcommand{\blind}{0}

\begin{document}

\def\spacingset#1{\renewcommand{\baselinestretch}%
  {#1}\small\normalsize} \spacingset{1}


\if0\blind { \title{\bf Model Selection using Multi-Objective
    Optimization} \author{Perry J. Williams \thanks{ The authors
      gratefully acknowledge the U.S. Geological Survey, Alaska
      Science Center, the Colorado State University Department of
      Fish, Wildlife, and Conservation Biology, and the Department of
      Statistics.  Joel Schmutz and Yu Wei provided comments that
      improved an earlier version of this manuscript. Any use of
      trade, firm, or product names is for descriptive purposes only
      and does not imply endorsement by the
      U.S. Government.}\hspace{.2cm}\\
    Department of Natural Resources and Environmental Science\\
    University of Nevada, Reno\\
    and \\
    William L. Kendall \\
    U.S. Geological Survey, Colorado Cooperative Fish and Wildlife
    Research Unit \\
    Department of Fish, Wildlife, and Conservation Biology \\
    Colorado State University \\
    and \\
    Mevin B. Hooten \\
    U.S. Geological Survey, Colorado Cooperative Fish and Wildlife
    Research Unit \\
    Department of Fish, Wildlife, and Conservation Biology \\
    Department of Statistics\\
    Colorado State University \\
  }

  \maketitle
} \fi

\if1\blind { \bigskip \bigskip \bigskip
  \begin{center}
    {\LARGE\bf Model Selection using Multi-Objective Optimization}
  \end{center}
  \medskip } \fi

\pagebreak
\begin{abstract}
  Choices in scientific research and management require balancing
  multiple, often competing objectives. \emph{Multiple-objective
    optimization} (MOO) provides a unifying framework for solving
  multiple objective problems. Model selection is a critical component
  to scientific inference and prediction and concerns balancing the
  competing objectives of model fit and model complexity. The tradeoff
  between model fit and model complexity provides a basis for
  describing the model-selection problem within the MOO framework. We
  discuss MOO and two strategies for solving the MOO problem; modeling
  preferences pre-optimization and post-optimization. Most model
  selection methods are consistent with solving MOO problems
  \emph{via} specification of preferences pre-optimization. We
  reconcile these methods within the MOO framework. We also consider
  model selection using post-optimization specification of
  preferences. That is, by first identifying Pareto optimal solutions,
  and then selecting among them. We demonstrate concepts with an
  ecological application of model selection using avian species
  richness data in the continental United States.

\end{abstract}

\noindent%
{\it Keywords:} competing models, decision theory, model selection,
multiple objectives, Pareto frontier, optimal solution \vfill

\spacingset{1.45} 

\section{INTRODUCTION}
The goal of modeling scientific processes varies from identifying the
important factors driving a system, to robust prediction into the
future or across space. Multiple competing models are considered in
most cases, each model based on hypotheses of spatial or temporal
structure in parameters, heterogeneity among individuals within a
population, or candidates for covariates influencing the process of
interest. Ultimately, one model from the candidate models, or a
composition of candidate models, is selected for inference or
prediction. Model selection is one of the most common problems in
scientific research, and numerous model-selection methods are
available \citep[e.g.,][]{akaike1973information, mallows1973some,
  schwarz1978estimating, gelfand1998model, burnham2002model,
  hooten2015guide}. Each model-selection method represents an approach
to balancing the bias due to missing important factors (model fit)
with imprecision due to overfitting the data (model complexity). Each
method represents a different \emph{a priori} weighting of the
relative importance of model fit and model complexity. While
guidelines exist, there is no consensus among statisticians on best
methods for this model selection process \citep{hooten2015guide}.

Multi-objective optimization (MOO) is a formal decision-theoretic
framework for optimizing problems with more than one objective
\citep{marler2004survey, williams2016methods, williams2017guide}. MOO
is commonly used in engineering, economics, and other fields for which
decisions must balance trade-offs between $\geq2$ competing objectives
\citep{marler2004survey}.  When a decision maker has competing
objectives, a solution that is optimal for one objective might not be
optimal for the other objective and a single solution that optimizes
multiple objectives does not exist. With competing objectives there
exists many (possibly infinite) solutions that might be considered
``optimal'' \citep[i.e., Pareto
optimal;][]{williams2017guide}. However, in most decision contexts, a
decision maker can only make one choice (e.g., which model to use to
predict into the future?).  To choose among solutions, a decision
maker must include their preferences among objectives to identify a
final solution. MOO provides a mathematical framework for quantifying
preferences for examining multi-objective problems.

The MOO framework is described generally as
\begin{align}
  \begin{split}
    & \boldsymbol{f}(\boldsymbol{\theta^*})
    =\text{optimum}_{\theta}\boldsymbol{f}(\boldsymbol{\theta}),
    \label{eq:moo}
  \end{split}
\end{align}
where
$\boldsymbol{f}(\boldsymbol{\theta})=(f_1(\boldsymbol{\theta}),...,f_k(\boldsymbol{\theta}))$,
such that $g_j(\boldsymbol{\theta}) \leq c_j,~j=1,2,...,J,$ and
$h_l(\boldsymbol{\theta}) = d_l,~~l=1,2,...,L$,
$f_i(\boldsymbol{\theta})$ represent the $k$ different, potentially
competing, objective functions, $\boldsymbol{f}(\boldsymbol{\theta})$
is a vector of the different objective functions, $g_j$ and $h_l$
represent $J$ inequality constraints and $L$ equality constraints,
respectively, and $\boldsymbol{\theta}$ is a vector of design
variables \citep{marler2004survey,cohon2013multiobjective}.

Pareto optimality is a concept of optimality used for eq. \ref{eq:moo}
when no value of $\boldsymbol{\theta}$ simultaneously optimizes all
functions $f_i$. A Pareto optimal solution for a minimization problem
is a solution $\boldsymbol{\theta^*} \in \boldsymbol{\Theta}$ for
which there is no other solution
$\boldsymbol{\theta}\in \boldsymbol{\Theta}$ such that both
$f(\boldsymbol{\theta}) \leq f(\boldsymbol{\theta^*})$, and
$f_i(\boldsymbol{\theta}) < f_i(\boldsymbol{\theta^*})$ for at least
one function $i$ \citep{deb2001multi,marler2004survey}. For decision
problems with competing objectives, there are many (potentially
infinite), Pareto optimal solutions. The set of solutions that are
Pareto optimal are known as the Pareto set (or Pareto frontier or
efficiency frontier). Each solution in a Pareto set has an implied set
of preferences for the objective functions $f_i$ \citep{deb2001multi,
  williams2016combining}. Thus, choosing among a set of Pareto optimal
solutions requires assuming (either implicitly or explicitly)
preferences for the objective functions $f_i$. Preferences among
objective functions can be specified pre- or post-optimization,
representing two separate strategies to solving eq. \ref{eq:moo}
\citep{williams2017guide}. When specifying preferences
pre-optimization, decision makers explicitly describe preferences of
objective functions and select the Pareto optimal solution associated
with their choice of preferences. When specifying preferences
post-optimization, decision makers first examine the set of Pareto
optimal solutions. Then the decision maker chooses the final Pareto
optimal solution based on the trade-offs observed among the set. The
choice implies decision-maker preferences.

One of the most common methods for incorporating preferences for $f_i$
into a decision problem pre-optimization, is the weighted-sum method
\citep{athan1996note,das1997closer,cohon2013multiobjective,
  williams2017guide}. The weighted-sum method is described by
\begin{align}
  f(\boldsymbol{\theta}) = \sum_{i=1}^k w_if_i(\boldsymbol{\theta}),
  \label{eq:linear}
\end{align}
for which the optimal solution is
\begin{align}
  \begin{split}
    &f(\boldsymbol{\theta}^*)=\text{optimize}_{\boldsymbol{\theta}}\sum_{i=1}^k
    w_if_i(\boldsymbol{\theta}).
    \label{eq:wsoptimum}
  \end{split}
\end{align}
The weights $w_i$ are chosen by the decision maker to reflect the
importance of each objective function $f_i$. The weighted-sum method
is a composition that results in a single objective function over
which to optimize. When optimizing one objective function, an
unequivocal optimal choice can be made.

We examine model selection within the MOO framework and demonstrate
that several methods commonly used for model selection in scientific
research are specific cases of the MOO problem solved using the
weighted-sum method with \emph{a priori} specification of
preferences. We examine concepts of the MOO framework, specifically
Pareto optimality, as it relates to several common model selection
methods. Finally, we examine the second strategy of MOO,
post-specification of preferences, and its application to the model
selection problem in scientific research. We demonstrate the concepts
presented using an example from the field of ecology involving
variable selection in a generalized linear regression model for avian
species richness data.

\section{MODEL SELECTION AS A MOO PROBLEM}
Methods for model selection typically consist of minimizing a weighted
sum of two functions, often described heuristically as a function for
model fit and a function for model complexity
\citep[e.g.,][p. 87]{burnham2002model}. That is, from eq. 2 we obtain
\begin{align}
  \begin{split}
    &f(\boldsymbol{\theta}^*)=\text{min}_{\boldsymbol{\theta}}\sum_{i=1}^2
    w_if_i(\boldsymbol{\theta}),
    \label{eq:common}
  \end{split}
\end{align}
where $\boldsymbol{\theta}^*$ represents the optimal solutions from
the set of design variables $\boldsymbol{\theta}$ (i.e., model
parameters), describing fit and complexity of any model, $w_i$ are
weights for the importance of the objectives associated with model fit
and complexity, and $f_i$ are functions that quantify the value of
model fit and complexity. Clearly, eq. \ref{eq:common} is a specific
form of the MOO problem defined in eq. \ref{eq:wsoptimum}. Theoretical
justification exists for choices of objective functions
$f_i(\boldsymbol{\theta})$ and their corresponding weights $w_i$
\citep{akaike1973information,mallows1973some,schwarz1978estimating,
  gelfand1998model,burnham2002model,link2006model,hooten2015guide}. Although
there is no consensus among statisticians on specific model selection
methods, most of the theoretical development related to model
selection can be described by two general functions for
$f_i$. Differences in model selection criteria are often the result of
different choices in weights. The most common objective function for
model fit is the negative log-likelihood of the data, given parameters
(i.e., the deviance). That is, if $f_1$ is the objective function
associated with model fit, it is described as
\begin{align}
  \begin{split}
    f_1(\boldsymbol{\theta}) =
    -\text{log}(L(\boldsymbol{\theta}|\boldsymbol{y})).
    \label{eq:f1}
  \end{split}
\end{align}
Although the deviance is the most common objective function for model
fit, others have been used. For example in Mallows' $C_p$,
$f_1(\boldsymbol{\theta})$ =
$\frac{\sum_{i=1}^n(y_i-\hat{\mu}_{\text{sub}})^2}{\sum_{i=1}^n(y_i-\hat{\mu}_{\text{full}})^2}-n$,
where $\hat{\mu}_{\text{sub}}$ equals the estimated mean of a
sub-model in consideration, $\hat{\mu}_{full}$ equals the estimated
mean of the full model in consideration, and $n$ equals the sample
size \citep{mallows1973some}.

\cite{hooten2015guide} summarize several objective functions for model
complexity using a function proportional to
\begin{align}
  &f_2(\boldsymbol{\theta}) =
    \sum_{j=1}^{p}|\theta_j-\mu_j|^{\gamma},
    \label{eq:f2}
\end{align}
known as the \emph{regulator}, \emph{regularizer}, or
\emph{penalty}. In eq. \ref{eq:f2}, $p$ represents the number of
parameters in the model, $\gamma$ is the degree of the norm; a
user-defined parameter that controls the relative penalty of the
distance between $\theta_j$ and $\mu_j$, $\theta_j$ are parameter
estimates for centered and scaled covariates, and $\mu_j$ is a
location parameter, often set to 0. Substituting the choices of
$f_1(\boldsymbol{\theta})$ and $f_2(\boldsymbol{\theta})$ from
eqs. \ref{eq:f1} and \ref{eq:f2} into eq. \ref{eq:common}, we obtain
the following multi-objective optimization problem
\begin{align}
  f(\boldsymbol{\theta})&=w_1f_1(\boldsymbol{\theta}) + w_2f_2(\boldsymbol{\theta}),\notag\\
                        &=w_1(-\text{log}(L(\boldsymbol{\theta};\mathbf{y})))
                          +w_2\sum_{j=1}^{p}|\theta_{j}-\mu_j|^{\gamma},
                          \label{eq:modselection}
\end{align}
with the objective of
$\text{min}_{\boldsymbol{\theta}}(f(\boldsymbol{\theta}))$.  Equation
\ref{eq:modselection} is the general function used in many model
selection methods including Akaike's information criterion (AIC), AIC
for small samples ($\text{AIC}_c$), quasi-AIC (QAIC), QAIC for small
samples (QAIC$_c$), Schwartz's information criterion (BIC), ridge
regression, LASSO (least absolute shrinkage and selection operator),
natural Bayesian shrinkage, and some forms of posterior predictive
loss \citep[Table 1; ][]{gelfand1998model,hooten2015guide}. Each of
the listed model selection methods result from specific choices of
$\boldsymbol{w}$ and $\gamma$, which we report in Table 1. For
example, let the weights be: $w_1=2$, $w_2=2$, and set $\gamma$ to
zero. With these weights, eq. \ref{eq:modselection} simplifies to
$-2\text{log}(L(\boldsymbol{\theta}|\boldsymbol{y})) + 2p$, or AIC
(Table 1).

Expressing model selection methods in terms of eq. \ref{eq:wsoptimum}
has an important result that links model selection to Pareto
optimality.  For positive weights $\boldsymbol{w}$, any solution to
eq. \ref{eq:wsoptimum} is a Pareto optimal solution
\citep{marler2010weighted}. Thus, any model selection method that can
be expressed in terms of eq. \ref{eq:modselection} (i.e., the methods
in Table 1) results in a solution that is Pareto optimal with respect
to the objectives of maximizing model fit and minimizing model
complexity.


\begin{table}[ht!]
  \centering
  \begin{tabular}{p{2cm} p{2cm} p{2.5cm} p{1cm} p{3cm}}
    \hline
    Model selection method & $w_1$ & $w_2$ & $\gamma$ & Note \\
    \hline
    AIC      & 2 & 2 & 0 & \\ [5ex]
    AIC$_c$  & 2 & $2(\frac{n}{n-p-1})$ & 0 & \\ [5ex]
    QAIC     & $\frac{2}{\hat{c}}$ & 2 & 0 & $\hat{c}=\chi^2/df$ \\ [5ex]
    QAIC$_c$ & $\frac{2}{\hat{c}}$ & $2(\frac{n}{n-p-1})$ & 0 & $\hat{c}=\chi^2/df$\\[5ex]
    BIC      & 2 & log($n$) & 0 & \\ [5ex]
    $^*$Ridge \\regression & 1 & User defined or estimated & 2 & Larger values of $w_2$ shrink $\boldsymbol{\beta}$ to 0.\\[5ex]
    $^*$LASSO & 1 & User defined or estimated & 1 & Larger values of $w_2$ shrink $\boldsymbol{\beta}$ to 0.\\
    \hline
  \end{tabular}
  \caption{Values of weights ($w_i$) and $\gamma$ for the
    multi-objective optimization problem of model selection described
    in eq. \ref{eq:modselection} for various model selection
    methods. The objective function for model fit is
    -log($L(\boldsymbol{\theta}|\boldsymbol{y})$), where
    $\boldsymbol{\theta}\equiv \boldsymbol{\beta}$; the objective
    function for model complexity is
    $\sum_{j=1}^p|\beta_j-\mu_j|^{\gamma},j=1,...,p$. AIC = Akaike's
    information criterion; AIC$_c$ = Second-order information
    criterion; QAIC = quasi-AIC; BIC = Schwartz information criterion;
    $n$ = sample size; $p$ = no. parameters in model. ($^*$) indicates
    objective function for model fit defined by:
    $\sum_{i=1}^n(y_i-\beta_0-\boldsymbol{x}'\boldsymbol{\beta})^2$. See
    \citet{burnham2002model} and \citet{hooten2015guide} for
    additional details. \vspace{5mm}}
\end{table}

\section{MODEL SELECTION USING POST-OPTIMIZATION SELECTION OF WEIGHTS}

Solving a MOO problem with competing objectives using
post-optimization specification of weights requires first identifying
as many Pareto optimal solutions as possible, then choosing among the
Pareto optimal solutions \citep{williams2017guide}. Pareto optimal
solutions for the objective functions in eqs. \ref{eq:f1} and
\ref{eq:f2} are models for which increasing the value of
eq. \ref{eq:f1} requires a decrease in the values in eq. \ref{eq:f2},
and \emph{vice versa}. One method for identifying Pareto optimal
solutions with two objective functions, each depending on
$\boldsymbol{\theta}$, is to plot the values of eqs. \ref{eq:f1} and
\ref{eq:f2} for each candidate model on opposing axes to identify the
Pareto frontier (e.g., Fig. \ref{fig:aic}). After the Pareto frontier
is identified, the decision maker can select the model based on the
trade-offs observed in the Pareto frontier. This is analogous to
\emph{best subset selection}, an active area of statistical research
\citep[e.g.,][]{hastie2009elements}. Thus, the selection of the final
model is made without explicitly choosing weights $\boldsymbol{w}$
associated with the model selection criteria listed in Table
1. However, if a choice from the Pareto frontier is also optimal with
respect to specific model-selection criterion, the weights of that
selection criterion are implied.

\section{EXAMPLE: AVIAN SPECIES RICHNESS IN THE U.S.}

Model selection is regularly used in the field of ecology to select
variables to include in linear and generalized linear regression
models. We examine the variable selection problem within a MOO
framework by considering avian species richness in the contiguous
U.S. as a function of state-level covariates. These data were
originally used to demonstrate model selection techniques in
\cite{hooten2017comparing}. As in \cite{hooten2017comparing}, we seek
to model the number of avian species $y_i$ ($i=1,\ldots,49$) counted
in each of the contiguous states in the U.S. and Washington D.C.,
based on covariate information $\textbf{x}_i$ collected in each
state. The covariate information includes the area of the state, the
average temperature, and the average precipitation. We modeled count
data using a Poisson distribution
\begin{align}
  y_i &\sim \text{Poisson}(\lambda_i), \notag
\end{align}
where $\lambda_i$ represents the mean species richness in each
state. We linked mean species richness to the covariate data using the
log link function
\begin{align}
  \text{log}(\lambda_i)&=\beta_0+\beta_1x_{1,i}+\ldots+\beta_px_{p,i}.
                         \label{eq:link}
\end{align}
We considered a total of 24 different models, representing different
linear and quadratic combinations of eq. \ref{eq:link}; each of the 24
candidate models are provided in Table \ref{tab:aic}. We used the
\texttt{glm} function in R statistical software version 3.3.2
\citep{r332} to fit the models to the data. Code to fit the models and
plot Fig. \ref{fig:aic} is provided in the Appendix.

\subsection{Model selection using AIC}

To conduct model selection for the avian species richness data, we
used the objective function in eq. \ref{eq:modselection} with values
of $\boldsymbol{w}\equiv 2$, and $\gamma=0$ (i.e., AIC). That is, for
a Poisson likelihood, the weighted objective function was
\begin{align}
  f(\boldsymbol{\beta}_m)&=2 \bigg(
                           \sum_{i=1}^n(\lambda_{i,m}-y_i\text{log}(\lambda_{i,m})+\text{log}(y_i!))
                           \bigg)
                           +2\sum_{j=1}^{p_m}|\beta_{j,m}|^{0},
                           \label{eq:linreg}
\end{align}
where $\boldsymbol{\beta}_m$ is the subset of parameters for model
$m=1,...,24$, $n$ is the sample size, and the term $\text{log}(y_i!)$
can be omitted because it is independent of $\boldsymbol{\beta_m}$,
and therefore, constant among models, provided the likelihood is not
changed.

The model from Table \ref{tab:aic} that minimized eq. \ref{eq:linreg}
(i.e., the AIC top model) included the intercept a linear area effect,
and a quadratic precipitation and temperature effect. All other model
fitting results are shown in Table \ref{tab:aic}.

\begin{landscape}
  \begin{longtable}{l c c c c}
    \captionsetup{width=9.5in}
    \caption{Model selection results from avian species richness
      data. AIC is Akaike's information criterion, $\Delta\text{AIC}=$
      is the difference in AIC compared to the top model. Asterisks
      ($^*$) indicate Pareto optimal models, $f_1(\boldsymbol{\beta})$
      and $f_2(\boldsymbol{\beta})$ are described in
      eq. \ref{eq:f1} and eq. \ref{eq:f2}, respectively.}\\
    \hline
    $\text{log}(\lambda_i)=$ &  $f(\boldsymbol{\beta})$   & $\Delta$AIC & $f_1(\boldsymbol{\beta})$ & $f_2(\boldsymbol{\beta})$ \\
    & (i.e., AIC)                &             & &   \\
    \hline
    $\beta_0$ & 741.1 & 229.8 & 369.6 & 1$^*$  \\
    $\beta_0 + \beta_1x_{\text{area},i}$ & 571.2 &59.9 & 283.6 & 2$^*$  \\
    $\beta_0 + \beta_1x_{\text{temp},i}$ & 669.2 & 157.9 & 332.6 & 2  \\
    $\beta_0 + \beta_1x_{\text{precip},i}$ & 706.1 &194.8 & 351.0 & 2  \\
    $\beta_0 + \beta_1x_{\text{area},i} + \beta_2x_{\text{temp},i}$ &
    526.7 & 15.4 & 260.3 & 3$^*$  \\
    $\beta_0 + \beta_1x_{\text{area},i} + \beta_2x_{\text{precip},i}$
    & 567.7& 56.4 & 280.9 & 3  \\
    $\beta_0 + \beta_1x_{\text{temp},i} + \beta_2x_{\text{precip},i}$
    & 536.7 & 25.5 & 265.4 & 3  \\
    $\beta_0 + \beta_1x_{\text{area},i} +
    \beta_2x_{\text{area},i}^2$ & 572.8 & 61.5 & 283.4 & 3  \\
    $\beta_0 + \beta_1x_{\text{temp},i} +
    \beta_2x_{\text{temp},i}^2$ & 668.0 & 156.7 & 331.0 & 3  \\
    $\beta_0 + \beta_1x_{\text{precip},i} +
    \beta_2x_{\text{precip},i}^2$ &704.9 &193.6 & 349.4 & 3  \\

    $\beta_0 + \beta_1x_{\text{area},i} + \beta_2x_{\text{temp},i}
    + \beta_3x_{\text{precip},i}$ & 515.3 & 4.0  & 253.7 & 4$^*$ \\
    $\beta_0 + \beta_1x_{\text{area},i} + \beta_2x_{\text{area},i}^2+
    \beta_3x_{\text{temp},i}$ & 524.3 &
    13.0 & 258.1 & 4 \\

    $\beta_0 + \beta_1x_{\text{area},i} + \beta_2x_{\text{area},i}^2+
    \beta_3x_{\text{precip},i}$ & 565.6 & 54.3 & 278.8 & 4
    \\
    $\beta_0 + \beta_1x_{\text{temp},i} + \beta_2x_{\text{temp},i}^2+
    \beta_3x_{\text{area},i}$ & 528.2 &
    16.9 & 260.1  & 4 \\
    $\beta_0 + \beta_1x_{\text{temp},i} + \beta_2x_{\text{temp},i}^2+
    \beta_3x_{\text{precip},i}$ & 535.5 &
    24.2  & 263.8 & 4 \\
    $\beta_0 + \beta_1x_{\text{precip},i} +
    \beta_2x_{\text{precip},i}^2+ \beta_3x_{\text{area},i}$ & 568.1 &
    56.8  & 280.0  & 4\\
    $\beta_0 + \beta_1x_{\text{precip},i} +
    \beta_2x_{\text{precip},i}^2+ \beta_3x_{\text{temp},i}$ & 524.2 &
    12.9  & 258.1 & 4\\

    $\beta_0 + \beta_1x_{\text{area},i} + \beta_2x_{\text{area},i}^2+
    \beta_3x_{\text{temp},i} +
    \beta_4x_{\text{temp},i}^2$ &525.7 & 14.4 & 257.8 & 5 \\

    $\beta_0 + \beta_1x_{\text{area},i} + \beta_2x_{\text{area},i}^2+
    \beta_3x_{\text{precip},i} +
    \beta_4x_{\text{precip},i}^2$ & 567.1 & 55.8 & 278.6 & 5 \\

    $\beta_0 + \beta_1x_{\text{temp},i} + \beta_2x_{\text{temp},i}^2+
    \beta_3x_{\text{precip},i} +
    \beta_4x_{\text{precip},i}^2$ &518.0 & 6.7 & 254.0 & 5  \\

    $\beta_0 + \beta_1x_{\text{area},i} + \beta_2x_{\text{area},i}^2+
    \beta_3x_{\text{precip},i} + \beta_4x_{\text{precip},i}^2+
    \beta_5x_{\text{temp},i} $ &512.2 & 0.9  & 250.1 & 6 \\

    $\beta_0 + \beta_1x_{\text{area},i} + \beta_2x_{\text{area},i}^2+
    \beta_3x_{\text{precip},i} + \beta_5x_{\text{temp},i} +
    \beta_6x_{\text{temp},i}^2$ &519.2 & 7.9 & 253.6 & 6  \\

    $\beta_0 + \beta_1x_{\text{area},i} + \beta_2x_{\text{precip},i} +
    \beta_3x_{\text{precip},i}^2+ \beta_4x_{\text{temp},i} +
    \beta_5x_{\text{temp},i}^2$ &511.3 & 0 & 249.6 & 6$^*$  \\
    $\beta_0 + \beta_1x_{\text{area},i} + \beta_2x_{\text{area},i}^2+
    \beta_3x_{\text{precip},i} + \beta_4x_{\text{precip},i}^2+
    \beta_5x_{\text{temp},i} + \beta_6x_{\text{temp},i}^2$ & 512.8 &
    1.5 & 249.4 & 7$^*$  \\
    \hline
    \label{tab:aic}
  \end{longtable}
\end{landscape}

Finally, a common practice for pre-specification of weights in MOO
problems in other applications includes examining the sensitivity of
the optimal choice relative to the selected weights
\citep{barron1988sensitivity,rios1990sensitivity}. An analogous
procedure in the model selection framework is to examine the optimal
solutions relative to different information criteria because different
criteria represent different objective weights (Table 1). The AIC,
AIC$_c$, and BIC criteria all resulted in the same top model
suggesting the optimal solution for these data was robust to several
different choices of weights.

\subsection{Model selection by examining Pareto optimal solutions}

\begin{center}
  \begin{figure}
    \doublespacing
    \includegraphics[width=.9\linewidth]{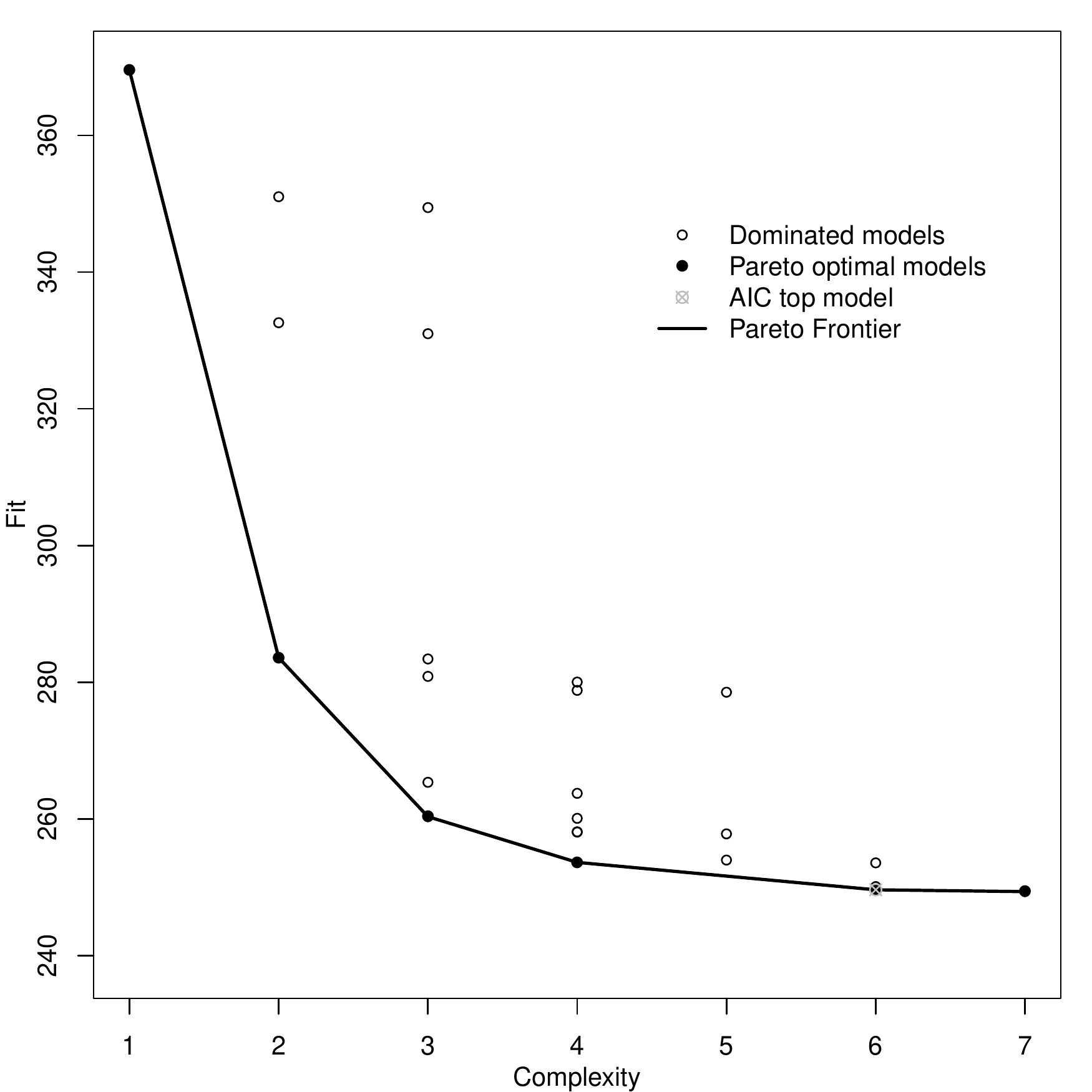}
    \caption{Model fit
      ($f_1(\boldsymbol{\theta})=-\text{log}(L(\boldsymbol{\theta}|\boldsymbol{y})$)
      vs. model complexity ($f_2(\boldsymbol{\theta})=$
      no. parameters) for each candidate model fit to avian species
      richness data.  Optimal solutions minimize fit (moving towards
      bottom of figure) and complexity (moving to the left of
      figure). The top model using $f(\boldsymbol{\theta})=\text{AIC}$
      was a Pareto optimal solution. Two \emph{candidate models}
      (i.e., $\Delta\text{AIC}<2$) were not Pareto optimal (i.e., they
      were dominated by another model).}
    \label{fig:aic}
  \end{figure}
\end{center}

Using avian species richness data, we examined model selection
\emph{via} specification of preferences post-optimization
(Fig. \ref{fig:aic}). That is, we identified Pareto optimal solutions
among the 24 models, and then considered potential methods for
selecting a model.  To identify Pareto optimal solutions, we used a
graphical approach and plotted the values $f_1$ and $f_2$ described in
eq. \ref{eq:linreg} for each model on opposing axes to identify
solutions along the Pareto frontier (Fig. \ref{fig:aic}). Identifying
Pareto optimal solutions does not require specifying $w_1$ or $w_2$,
and therefore does not require adhering to an information
criterion. The Pareto optimal set included 6 models; one model for
each number of parameters $1,\ldots,7$, except $p=5$, where both model
fit and complexity could be simultaneously improved by using the top
model containing four parameters. Each Pareto solution represented the
model that minimized eq. \ref{eq:linreg} among all models with the
same number of parameters. There were 17 dominated models (i.e.,
models that were not Pareto optimal; Fig. \ref{fig:aic}). The AIC top
model was a Pareto optimal solution; this was expected because AIC
(and other information criteria) is a specific formulation of the
weighted-sum method and is therefore \emph{sufficient} for Pareto
optimality \citep[][]{marler2010weighted}. Each of the Pareto
solutions correspond to a specific set of weights in eq.
\ref{eq:modselection}.

Given the information on Pareto optimal solutions in
Fig. \ref{fig:aic}, selecting a final model for inference can proceed
in many ways, depending on the application and the nature of the
parameters under consideration. A decision maker can use the
information on Pareto optimal solutions to view trade-offs of fit
gained by adding (or subtracting) additional parameters from the
model, and choose a Pareto optimal solution with trade-offs acceptable
to the decision maker.  Some parameters might be associated with
covariates for which annual data are difficult, expensive, or
impossible to collect. The trade-offs in terms of model fit can be
assessed relative to the expense of collecting additional data for
these parameters. If the increase in model fit from the Pareto optimal
solution that requires the additional (expensive) covariate data does
not justify the additional expense, another Pareto optimal solution
may be preferred.

Another approach is to examine the curvature of the Pareto
frontier. An elbow shape (e.g, Fig. \ref{fig:aic}: $p=3$) can be
identified, where increasing the number of parameters has diminishing
marginal returns in terms of $f_1$, and decreasing the parameter size
has a large affect on $f_1$.  In the avian species richness data, the
largest improvement in model fit, per parameter added, was adding area
to the null model (an 86 unit improvement to fit;
Fig. \ref{fig:aic}). Subsequent parameter additions showed diminishing
marginal returns in model fit; the second biggest improvement in model
fit was adding precipitation to the area model (23 units), followed by
adding temperature to the area + precipitation model (7 units). No
models with five parameters occurred on the Pareto frontier.

Another approach is to compare the trade-offs to biological
significance of the parameters involved and the need to make inference
on those parameters. For example, if a parameter is required to inform
a management decision, such as survival rates for harvest decisions, a
decision maker would prefer to choose a Pareto optimal solution that
included survival rates. Another approach might be to choose a Pareto
optimal solution such that the maximum number of parameters is
constrained by the amount of data. For example, if an investigator
wishes to constrain the number of parameters in the model such that
$p<\frac{n}{15}$, the investigator could select the Pareto optimal
solution that maximized model fit within the constrained set. In the
avian species richness data, with $n=49$, this would suggest choosing
the Pareto optimal model with three parameters (with log linear
predictor $\beta_0+\beta_1x_{area,t}+\beta_2x_{temp,t}$; Table
\ref{tab:aic}).

Finally, models that are optimal in terms of model selection criteria
could be highlighted as reference points on the Pareto frontier to
guide decisions on the final model choice.  Ultimately, the use of the
Pareto frontier is that it provides visual information on the
trade-offs of the objectives of the decision maker; in this case,
maximizing model fit and minimizing complexity.

\section{DISCUSSION}

The explicit application of multi-objective optimization to model
selection using the objective functions defined in eqs. \ref{eq:f1}
and \ref{eq:f2} ties several important properties of MOO to common
methods used in scientific research to select a model. First, many
different model selection methods are special cases of the
weighted-sum method; each method representing different objective
weights. This provides a unifying framework to quantitatively and
visually compare model-selection methods based on different
theoretical foundations.  Practitioners of multi-objective
optimization in operations research or other decision-theoretic fields
usually recommend sensitivity analyses of the resulting decisions
given the choice of objective weights \citep{keeney1993decisions,
  williams2017guide}. A sensitivity analysis for the model selection
problem consists of evaluating multiple model selection criteria
(representing different objective weights) to examine the robustness
of the solution to the choice of criterion. Many practitioners argue
against this approach, suggesting that a criterion should be selected
based on its theoretical motivation \citep[e.g., AIC is asymptotically
efficient; BIC is consistent,][]{aho2014model}. Others view a specific
information criterion as one line of evidence to assist in a decision
and report different criteria side-by-side
\citep[e.g.,][]{araujo2007importance, parviainen2008modelling}. The
former appears to be the dominant paradigm in ecological research,
whereas the latter is common in other fields. Second, many model
selection methods result in Pareto optimal solutions because they are
specific formulations of eq. \ref{eq:linear}, which is sufficient for
Pareto optimality. Thus, there is a decision-theoretic basis for model
selection methods that can be expressed in the form of
eq. \ref{eq:modselection} in terms of optimality criteria.

Although we described the model selection problem heuristically in
terms of maximizing model fit and minimizing model complexity, we
could have replaced model fit with predictive ability as the objective
of interest. Predictive ability is the most commonly sought model
characteristic for model selection, and many information criteria and
other model selection methods were developed to optimize predictive
ability \citep{akaike1973information, stone1977asymptotic,
  gelfand1998model, hoeting1999bayesian, burnham2002model,
  hooten2015guide}. Many information criteria have weights and
penalties that serve as bias corrections for optimization in terms of
predictive ability \citep{konishi1996generalised}. That is, many
information criteria are based on bias-corrected log likelihoods, for
which the model complexity is a correction factor to remove asymptotic
bias of the log likelihood of a fitted model
\citep{konishi1996generalised}. The MOO problem in terms of maximizing
predictive ability and accounting for model bias is similar in spirit
to the MOO problem of maximizing model fit while minimizing model
complexity. 

Model selection by examining trade-offs of fit and complexity
post-optimization has been used in several other applications. Users
of Mallows' C$_p$ often conduct similar investigations
\citep{mallows1973some}. \cite{freitas2004critical} examined Pareto
optimality in the related question comparing prediction and simplicity
for data
mining. Viewing each model's trade-offs, in terms of objectives, provides a
visual assessment of the model selection problem, a potentially useful
tool for ultimately choosing a model for inference or prediction. As
is the case with any multi-objective optimization problem, the
additional flexibility in model choice based on post-optimization
specification of preferences could be viewed as either a positive or
negative trait, depending on how an investigator values the order for
which preferences are specified. Specifying preferences
pre-optimization for the model selection problem benefits from being
objective in the sense that a decision maker chooses how to weigh
their specific objective functions without being influenced by how
weights will alter the outcome of optimization. Specifying preferences
post-optimization has the added flexibility of choosing a Pareto
optimal solution that provides the best trade-offs for context
dependent decision problems.

\bigskip
\begin{center}
  {\large\bf SUPPLEMENTARY MATERIAL}
\end{center}

\begin{description}

\item[Appendix:] R statistical software script to fit models described
  in Table 2 to avian species richness data, and calculate and plot
  values for eqs. 5 and 6 shown in Fig. 1.

\end{description}

\bibliographystyle{ecology} \bibliography{references}
\end{document}